\documentclass[rmp,11pt,onecolumn,floatfix,showkeys,byrevtex]{revtex4}

\usepackage{graphicx}
\usepackage{bm}
\usepackage{url}

\begin{document}

\title{Statistical Physics Perspective on Economic Inequality}

\author{Victor M.~Yakovenko}

\affiliation{JQI, Department of Physics, University of
  Maryland, College Park, Maryland 20742, USA}

\date{2 July 2023}


\thanks{ Professor of Physics; Ph.D.\ in Physics from the Landau Institute for Theoretical Physics in 1987 \\ 
ORCID \url{https://orcid.org/0000-0003-3754-1794} \\
Google Scholar \url{https://scholar.google.com/citations?user=pEnxwCMAAAAJ} \\
Web page \url{https://physics.umd.edu/~yakovenk/econophysics/} \\
E-mail \url{yakovenk@umd.edu} }

\begin{abstract}

This article is a supplement to my main contribution to the \textit{Routledge Handbook of Complexity Economics} (2023).  On the basis of three recent papers, it presents an unconventional perspective on economic inequality from a statistical physics point of view.  One section demonstrates empirical evidence for the exponential distribution of income in 67 countries around the world.  The exponential distribution was not familiar to mainstream economists until it was introduced by physicists by analogy with the Boltzmann-Gibbs distribution of energy and subsequently confirmed in empirical data for many countries.  Another section reviews the two-class structure of income distribution in the USA.  While the exponential law describes the majority of population (the lower class), the top tail of income distribution (the upper class) is characterized by the Pareto power law, and there is no clearly defined middle class in between.  As a result, the whole distribution can be very well fitted by using only three parameters.  Historical evolution of these parameters and inequality trends are analyzed from 1983 to 2018.  Finally, global inequality in energy consumption and CO$_2$ emissions per capita is studied using the empirical data from 1980 to 2017.  Global inequality, as measured by the Gini coefficient $G$, has been decreasing until around 2010, but then saturated at the level $G=0.5$.  The saturation at this level was theoretically predicted on the basis of the maximal entropy principle, well before the slowdown of the global inequality decrease became visible in the data.  This effect is attributed to accelerated mixing of the world economy due to globalization, which brings it to the state of maximal entropy and thus results in global economic stagnation.  This observation has profound consequences for social and geopolitical stability and the efforts to deal with the climate change.\\

\textsf{\normalsize ``Money, it's a gas.'' Pink Floyd, 
\textit{Dark Side of the Moon} (1973) }

\end{abstract}

\maketitle


\section{Introduction}
\label{Sec:Intro}

This article is a concise follow-up to my paper \cite{Yakovenko-2016} reproduced in this \textit{Routledge Handbook of Complexity Economics} (2023).  The update is based on three papers \cite{Tao-2019,Semieniuk-2020,Ludwig-2022} published after \cite{Yakovenko-2016} and summarized in three sections below.  All papers are available at \url{https://physics.umd.edu/~yakovenk/econophysics/}.

By analogy with the Boltzmann-Gibbs distribution\footnote{\url{https://en.wikipedia.org/wiki/Boltzmann_distribution}} of energy in statistical physics, \textcite{Dragulescu-2000} proposed that the stationary probability distribution of money in an ensemble of interacting economic agents follows an exponential law.  They performed agent-based modeling\footnote{\url{https://physics.umd.edu/~yakovenk/econophysics/animation.html}} to illustrate how the exponential Boltzmann-Gibbs distribution of money develops from an initially equal distribution due to monetary transactions between the agents, which are analogous to collisions between molecules in a gas.  The same computer simulations also shows that the entropy of the system increases and saturates at its maximal value in statistical equilibrium, in agreement with the second law of thermodynamics.

\textcite{Rosser-2016} highlighted distinction between two flavors of entropy: ``ontological'' and ``metaphorical.''  The former is the thermodynamic entropy introduced by Carnot, Clausius, and Boltzmann to physics in the 19th century.  It governs flow of energy in physical systems via the second law of thermodynamics.  The latter is the informational entropy introduced by Shannon and von Neumann in the 20th century as a measure of combinatorial complexity in any statistical ensemble, not limited to a physical ensemble.  The statistical physics perspective presented below is guided by application of the second, generalized concept of entropy to an ensemble of economic agents, either at national or global scale.  This perspective is the opposite to the representative-agent approach of traditional economics, which, by construction, reduces an ensemble to a single agent and thus ignores statistical and heterogeneous aspects of the economy, such as entropy and inequality \cite{Yakovenko-2022}.

\section{Exponential distribution of income in 67 countries}
\label{Sec:67countries}

\begin{figure}[b]
\includegraphics[width=0.49\linewidth]{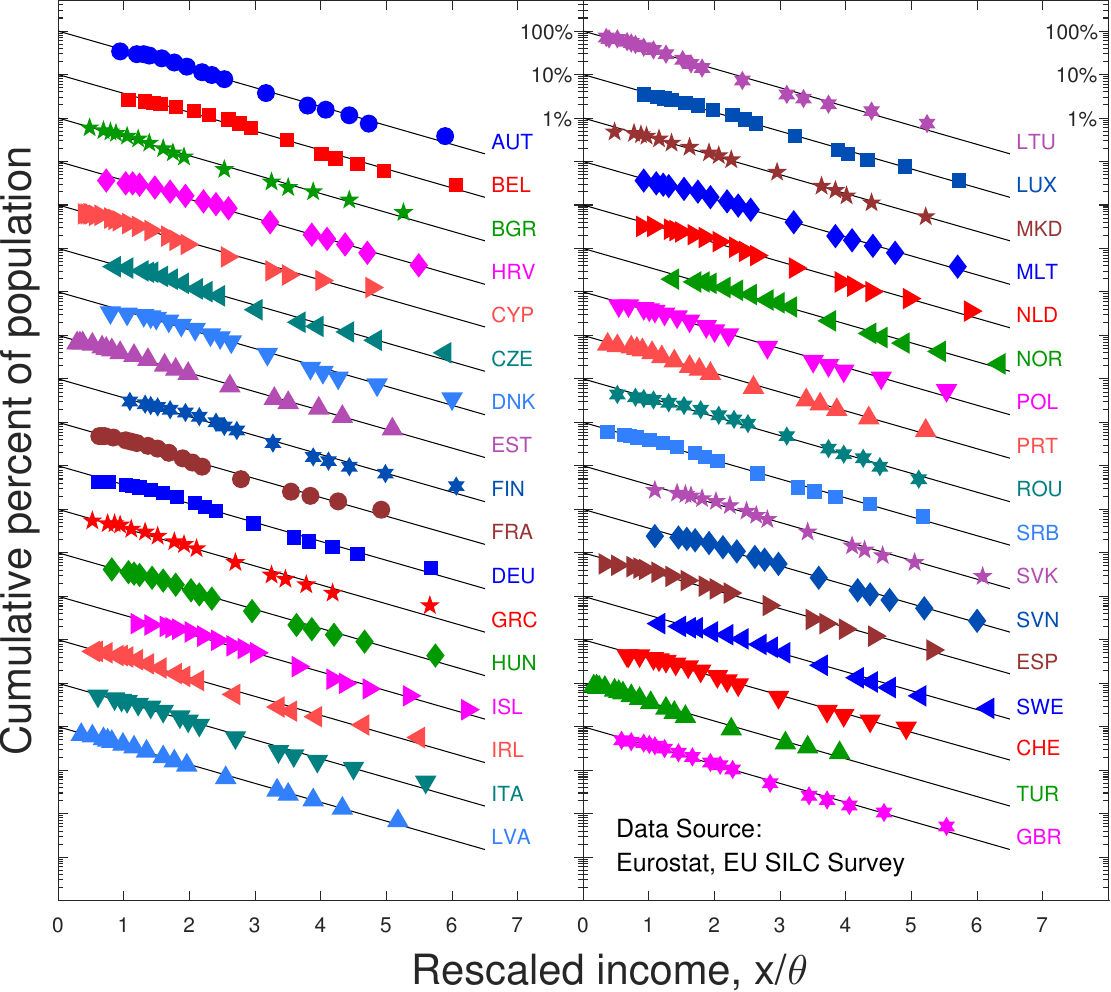} \hfill
\includegraphics[width=0.49\linewidth]{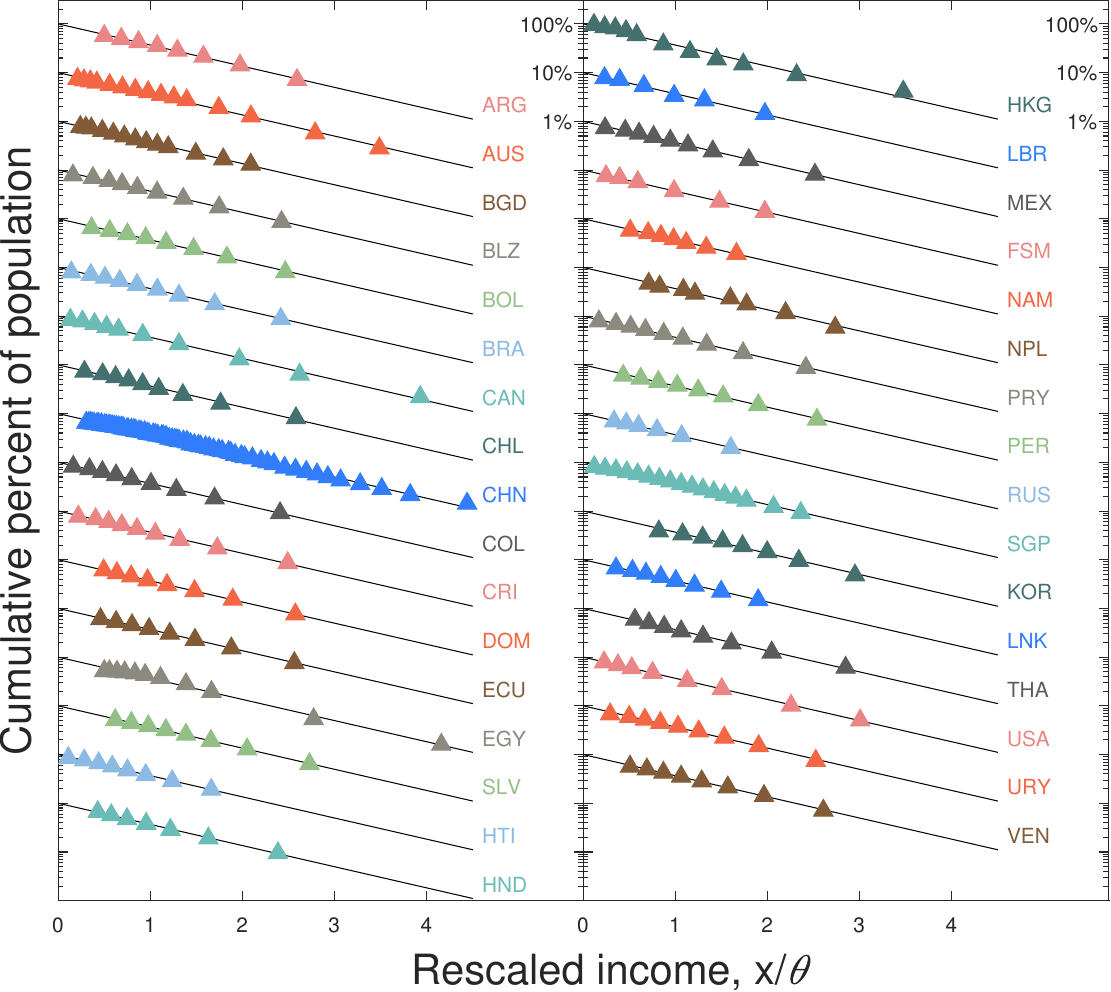}
\caption{Cumulative probability distribution in logarithmic scale on the vertical axis versus normalized income in linear scale on the horizontal axis for the European Union and its neighboring  countries in 2014 (left panel) and for select non-EU countries for various years (right panel).
[Reproduced from \textcite{Tao-2019}] }
\label{Fig:67countries}
\end{figure}

In their next paper, \textcite{Dragulescu-2001a} found empirical evidence for the exponential distribution of income in the USA by analyzing the data from the U.S.\ Census Bureau.  Further analysis of the data from the Internal Revenue Service (IRS, the U.S.\ tax agency) confirmed the exponential shape for the lower part of income distribution, where the overwhelming majority of population belongs, whereas the tail of income distribution for the top few percent of population was found to follow the Pareto power law \cite{Dragulescu-2001b,Dragulescu-2003}.  The observation that income distribution for the majority of population is described by the exponential law was virtually unknown to the economists and statisticians, but was eventually recognized in a special issue \textit{The Science of Inequality} of the \textit{Science} magazine in a one-page article by \textcite{Cho-2014}.

Over time, the exponential distribution of income was found in numerous papers for different countries, e.g., by \textcite{Banerjee-2006} for Australia.  A collaborative team with Chinese economists and data scientists \cite{Tao-2019} studied income distribution for 67 countries around the world (where it was possible to obtain reasonably reliable data).  The results are shown in Fig.~\ref{Fig:67countries} for the European Union (EU) and its neighboring  countries in the left panel and for select non-EU countries in the right panel.  The graphs are plotted in log-linear scale, where normalized income on the horizontal axis is in linear scale, whereas cumulative probability on the vertical axis is in logarithmic scale.  Data points following straight lines in log-linear scale confirm that the middle part of income distribution for these countries is, indeed, exponential.  The upper tail deviates from the exponential shape because of the Pareto law, but is not well sampled in the survey data, so it is truncated.  Income distribution at very low income may also deviate from the exponential shape because of welfare policies varying from country to country.  These low-end deviations are also truncated in Fig.~\ref{Fig:67countries}.  Overall, Fig.~\ref{Fig:67countries} demonstrates overwhelming empirical evidence that that the low and middle part of income distribution follows a universal exponential law for many countries.  For some countries, the exponential regime extents over a particularly wide range, e.g.,\ for China (CHN) in the right panel of Fig.~\ref{Fig:67countries}.

Besides the empirical analysis, \textcite{Tao-2019} also reformulated the arguments in favor of the exponential distribution from statistical physics to the language of economics using the standard Arrow-Debreu’s general equilibrium model, combined with Rawls’ fairness principle for free competition.

\section{Two-class structure of income distribution}
\label{Sec:2class}

\begin{figure}[b]
\includegraphics[width=0.525\linewidth]{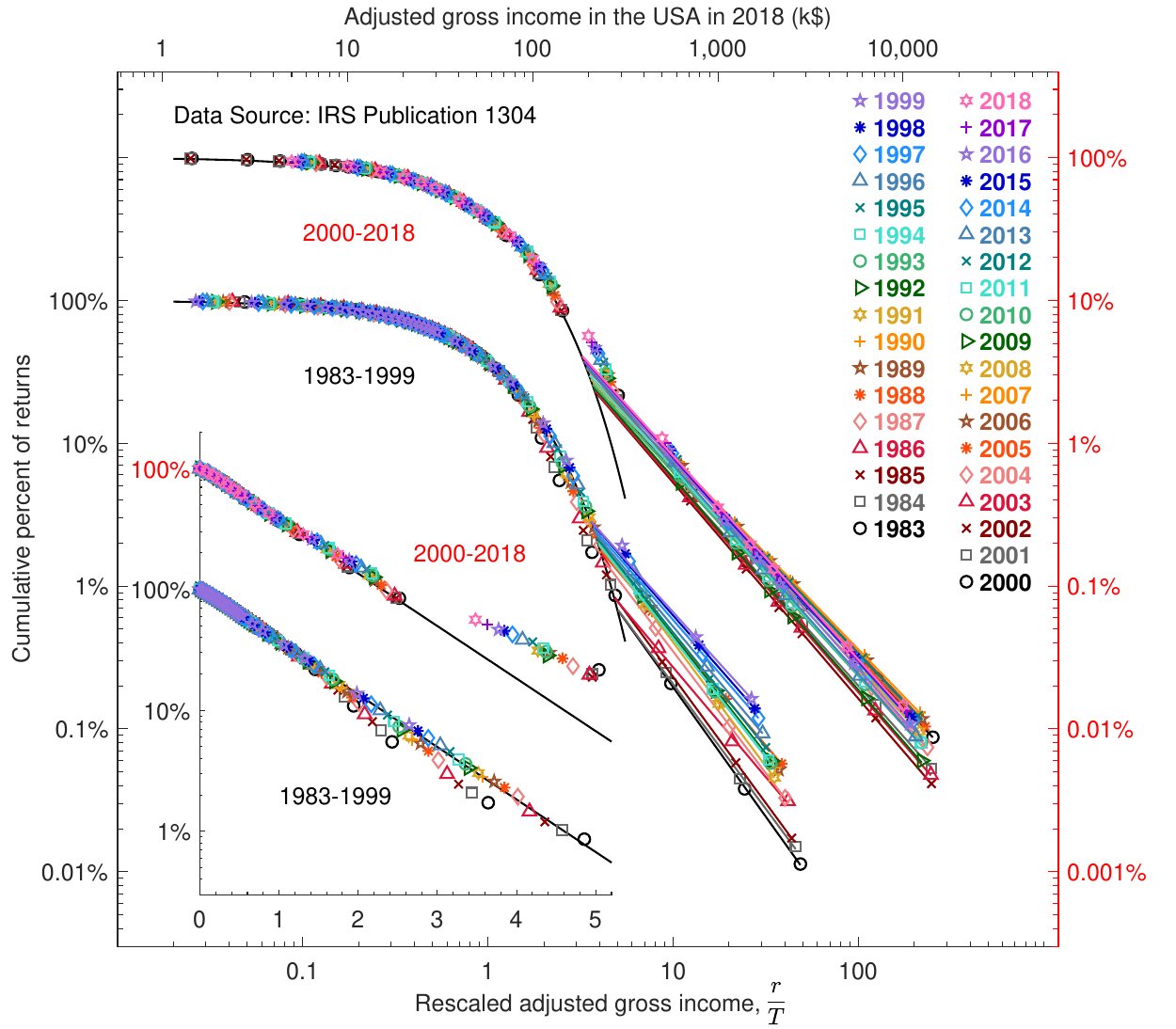} \hfill
\includegraphics[width=0.465\linewidth]{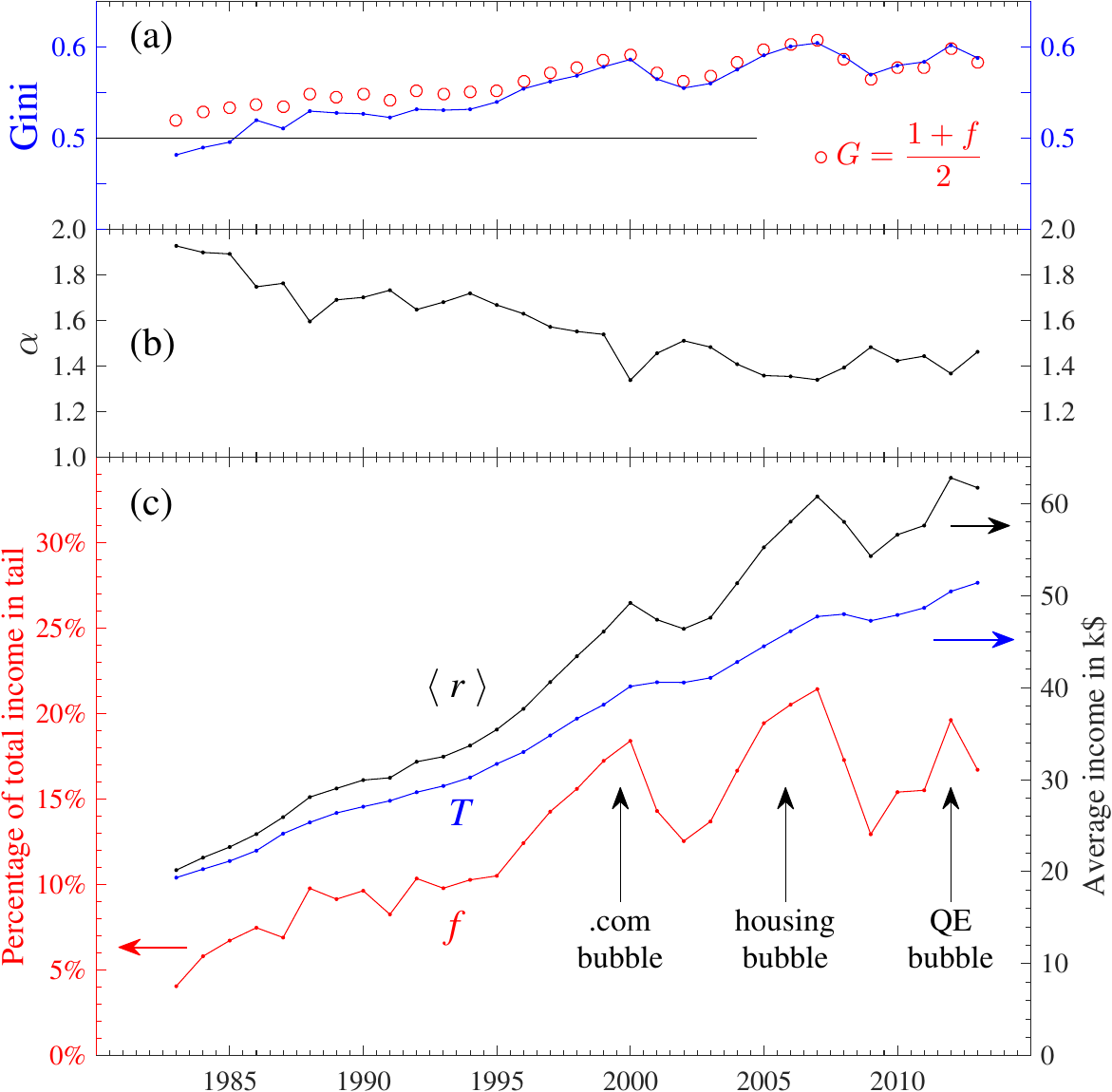}
\caption{Left panel: Cumulative distribution function of rescaled income in the USA for 1983--2018 in log-log scale (main panel) and log-linear scale (inset) [reproduced from \textcite{Ludwig-2022}].  
Right panel: (a) The Gini coefficient $G$ for 1983--2013 (connected line), compared with the formula $G=(1+f)/2$ (open circles).  (b) The exponent $\alpha$ of the power-law tail.  (c) The mean income $\langle r\rangle$ of the whole system, the mean income $T$ of the lower class (the temperature of the exponential part), and the fraction of income $f=1-T/\langle r\rangle$ going to the upper class.
[Reproduced from \textcite{Yakovenko-2016}] }
\label{Fig:2class}
\end{figure}

The papers reviewed in Sec.~\ref{Sec:67countries} considered income distribution for a given year.  \textcite{Silva-2005} extended this analysis to the multiyear period of 1983--2001 for the USA.  In subsequent papers \cite{Banerjee-2010,Yakovenko-2016} the range of years was expanded even further.  The latest paper \cite{Ludwig-2022} studied the two-class income distribution in the USA for 36 years from 1983 to 2018.  The cumulative distribution function versus rescaled income is shown for these years in the left panel of Fig.~\ref{Fig:2class}.  It clearly shows a two-class structure of income distribution.  The inset shows that the lower part of the distribution (about 96\% of population in 2018) is well described by an exponential function, as indicated by the straight lines in log-linear scale.  On the other hand, the main panel shows that the upper tail (about 4\% of population in 2018) is well described by a power law, as indicated by the straight lines in log-log scale.  Importantly, the data reveal only two classes: upper and lower, whereas middle class does not exist.  In fact, there is no commonly accepted definition of the ``middle class'' in the literature, and each author who writes about it invents their own definition.

Mathematically, the two-class distribution can be derived as a (quasi)stationary solution of the Fokker-Planck equation\footnote{\url{https://en.wikipedia.org/wiki/Fokker–Planck_equation}} (also known as the forward Kolmogorov equation in mathematics) for stochastic dynamics of income with coexisting additive and multiplicative components \cite{Silva-2005,Banerjee-2010,Yakovenko-Rosser-2009}.  The stationary distribution interpolates between an exponential law at the low end and a power law at the high end \cite{Ludwig-2022} and is known as the Pearson Type IV distribution \cite{Pearson-1895}.

Using the two-class decomposition, income distribution can be very well fitted by using only three parameters: the mean income $T$ of the exponential bulk (which is analogous to temperature in physics), the exponent $\alpha$ of the power-law tail, and the crossover income $r_*$ separating the lower and upper classes.  Historical evolution of the first two parameters is shown in the right panel of Fig.~\ref{Fig:2class}.  Panel (b) shows that the Pareto exponent $\alpha$ has been decreasing since 1983 (albeit flattening after 2000), which indicates ``fattening'' of the upper tail, i.e.,\ rich getting richer.  

Panel (c) of Fig.~\ref{Fig:2class} shows the income temperature $T$ in comparison with the mean income $\langle r\rangle$ of the whole distribution.  The normalized difference between these two parameters represents the share of income $f=(\langle r\rangle-T)/\langle r\rangle$ going to the upper class.  The spikes on the red curve for $f$ in Panel (c) indicate sharp increases in inequality due to the enhanced share of income going to the upper class.  The first spike coincides with the .com bubble in stock market, the second spike with the housing bubble, and the third spike with the Quantitative Easing (QE) pursued by the Federal Reserve.  Clearly, inequality peaks during speculative bubbles in the financial markets.  The third spike gives direct evidence that the bailout of the financial system by the Fed resulted in increase of inequality.  The parameter $f$ has increased even further in subsequent years \cite{Ludwig-2022}.  

Panel (a) of Fig.~\ref{Fig:2class} shows the Gini coefficient $G$ in comparison with the theoretical formula $G=(1+f)/2$ derived from the two-class decomposition by \textcite{Silva-2005}.  A very good agreement after 1995 indicates that the historical dynamics of the Gini coefficient is completely determined by the share of income $f$ going to the upper ``superthermal'' class, whereas relative inequality within the lower ``thermal'' class remains essentially constant over the span of 36 years \cite{Ludwig-2022}.  Furthermore, 
\textcite{Shaikh-2023} proposed a metric for ranking of countries, called the Vast Majority Income, on the basis of this formula for the Gini coefficient.

\textcite{Ludwig-2022} also studied the shares of tax revenue coming from the lower and upper classes, according to the IRS data.  By 2018, the income share of the top 1\% of the population has increased to 21\%, which is almost twice the total income share of 12\% going the bottom half (50\%) of the population.  At the same time, the tax share paid by the bottom 50\% of the population has decreased to 3\% and became almost negligible, because their total income share is so low. In contrast, the tax share of the top 1\% of the population has increased to 40\%.  Thus, the majority of tax revenue now comes from the top few percent of the population, where most of income is concentrated.

Not only does the upper-class income share increase, but the fraction of the population belonging to the upper class increases too \cite{Ludwig-2022}.  In a relative sense, the upper-class population expands (while still remaining a small fraction at 4\%), while the lower-class population shrinks (while still remaining the overwhelming majority).  \textcite{Ludwig-2022} speculated that it is due to digitization of the economy in the last 40 years. There was a rapid proliferation of personal computers during the 1980s, followed by the spread of the Internet and the World Wide Web in the 1990s, and then by ubiquitous personal mobile devices and a shift to cloud computing at the present time. This transformation enabled the creation and relatively easy scaling-up of digital platforms for highly non-local business operations. In the past, many businesses were local: taxi companies, book stores, video rental stores, etc.  Now they are largely displaced by a small number of national and global network platforms, such as Uber and Lyft for riding, Amazon for books initially and then for all kinds of goods, Netflix for DVD rentals and video streaming, etc. The founders and owners of such network platforms become super-rich, because these platforms serve a huge number of customers, in contrast to the old-fashioned local businesses. Thus the growth of the upper class may be a reflection of the ongoing transformation of business network topology from local clusters to highly connected superclusters and global hubs.

\section{Global Inequality in Energy Consumption and CO$_2$ Emissions}
\label{Sec:Global}

\begin{figure}[b]
\includegraphics[width=0.49\linewidth]{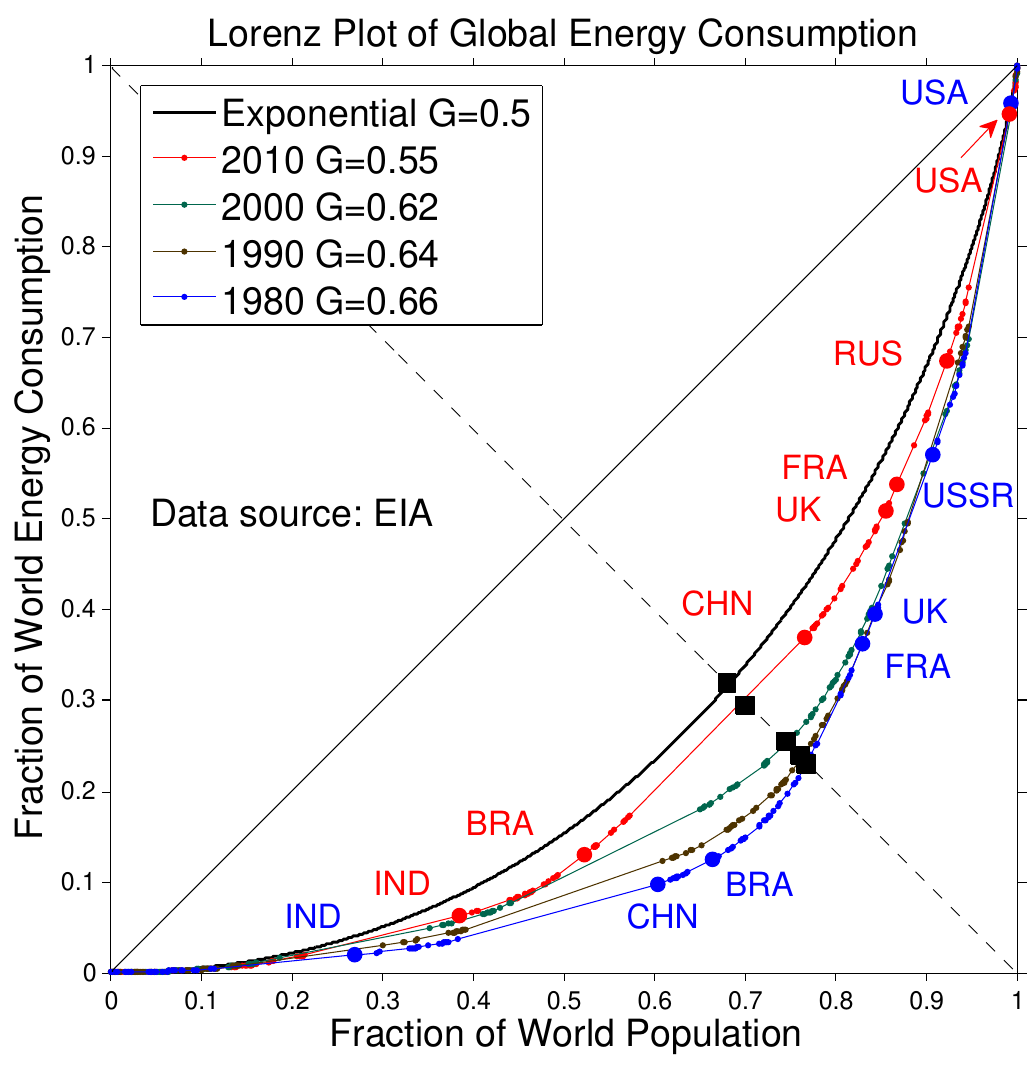} \hfill
\includegraphics[width=0.50\linewidth]{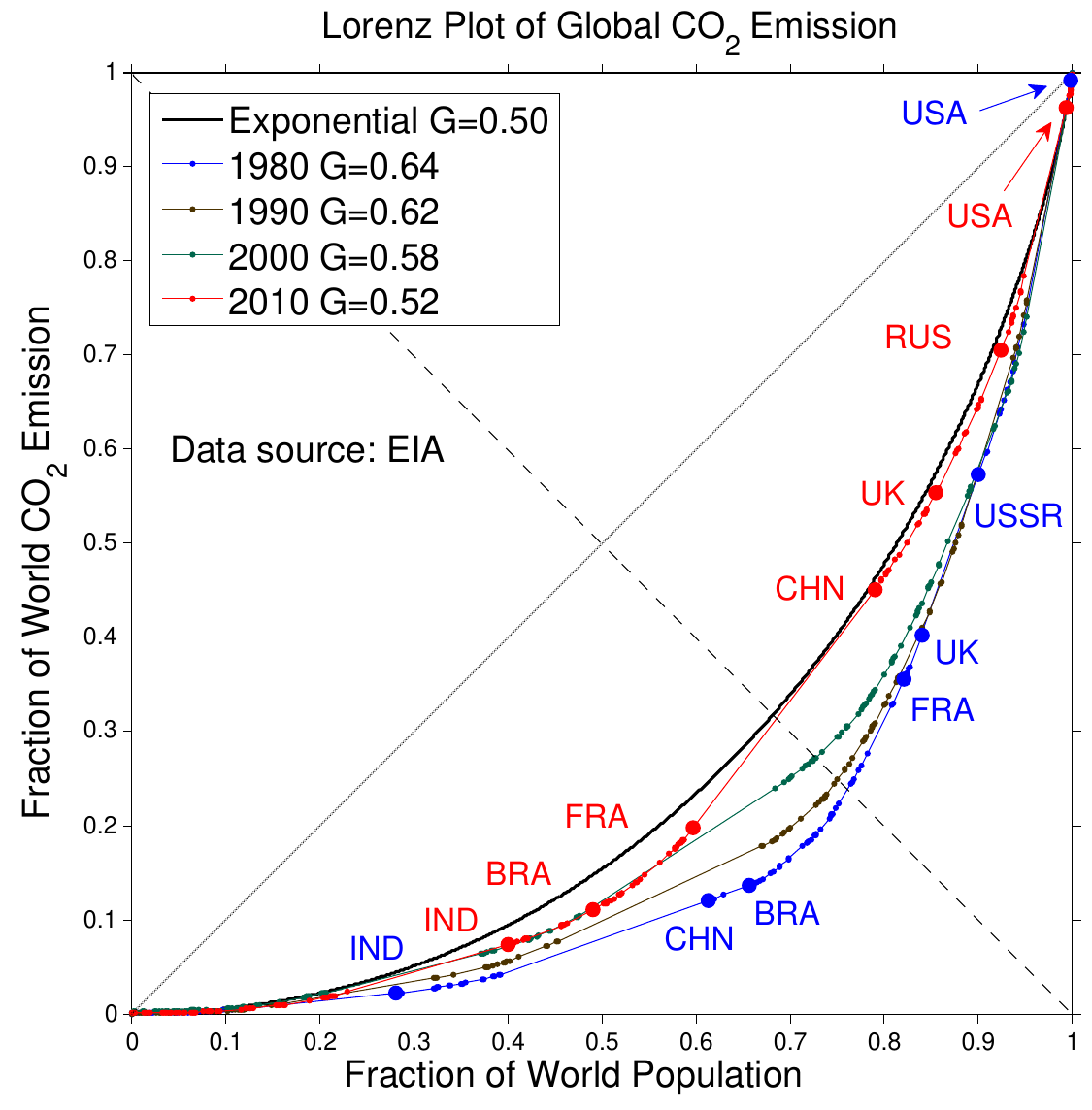}
\caption{Lorenz plots of the global energy consumption (left panel) and $\rm CO_2$ emissions (right panel) per capita for 1980, 1990, 2000, and 2010 (colored curves), compared with an exponential distribution (black curve). 
[Reproduced from \textcite{Lawrence-2013}] }
\label{Fig:Lorenz}
\end{figure}

Sections \ref{Sec:67countries} and \ref{Sec:2class} focus on income inequality within a given country.  However, there is also global inequality between rich and poor countries around the world.  Studying global monetary inequality is complicated by different currencies, whose nominal conversion rates are not particularly representative.  One way around this problem is to use the purchasing power parity\footnote{\url{https://en.wikipedia.org/wiki/Purchasing_power_parity}} \cite{Milanovic-2012}.  \textcite{Lawrence-2013} took another approach and investigated global inequality in energy consumption and CO$_2$ emissions per capita using the data from the U.S. Energy Information Administration (EIA) for 1980--2010.  Energy consumption is a physical measure of inequality in standards of living, whereas  CO$_2$ emissions are closely correlated with it, because most energy is produced from fossil fuels.

The corresponding Lorenz curves are shown in Fig.~\ref{Fig:Lorenz}.  Computer animation of the time evolution of these Lorenz curves is also available.\footnote{\url{https://physics.umd.edu/~yakovenk/econophysics/global.html}}  In both graphs, small circles indication various countries, whereas big circles indicate some labeled countries.  Not surprisingly, the Lorenz curves for energy consumption and CO$_2$ emissions look quite similar.

The Lorenz curves in 1980 exhibit a sharp slope change in the middle, which separates two groups of countries.  One group in the top-right sector has high slope, indicating high energy consumption and CO$_2$ emissions per capita.  This group consists of ``developed'' countries, mostly in North America and Europe.  Another group in the bottom-left sector has low slope, indicating low energy consumption and CO$_2$ emissions per capita.  This group consists of ``developing'' countries, such as China, India, and Brazil.

In the subsequent years since 1980, the Lorenz curves move up, indicating that global inequality \textit{decreases}.  By 2010, the cusp in the middle of the Lorenz curves has smoothed out.  Now there is no sharp boundary between ``developed'' and ``developing'' countries anymore. This is largely due to China moving to the middle of the curve between these two groups of countries.

\begin{figure}[b]
\includegraphics[width=0.50\linewidth]{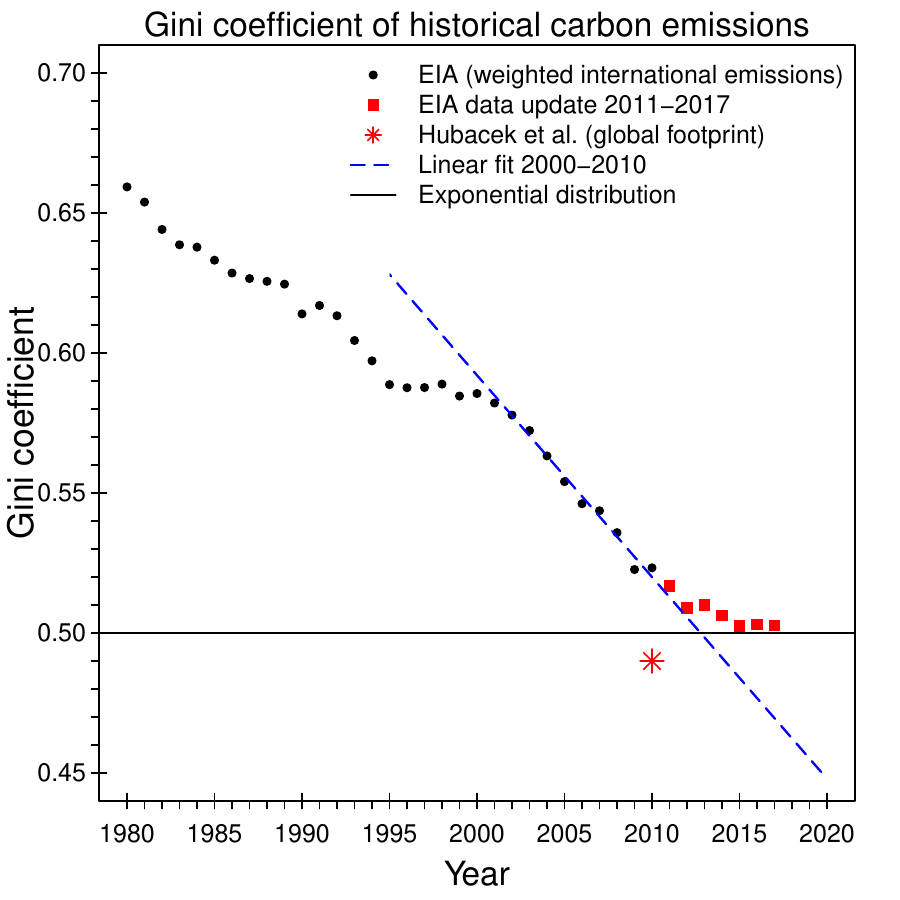}
\caption{Historical evolution of the global Gini coefficient for CO$_2$ emissions per capita.  Black circles show the data for 1980--2000 as published by \textcite{Lawrence-2013}, whereas red squares show the new EIA data for 2011--2017.  The blue dashed line is a linear extrapolation from 2000 to 2010.  The horizontal line at $G=0.5$ corresponds to a theoretical exponential distribution.
[Reproduced from \textcite{Semieniuk-2020}] }
\label{Fig:Gini}
\end{figure}

The black curve in Fig.~\ref{Fig:Lorenz} represents the (analytically) calculated Lorenz graph for an exponential distribution.  In statistical physics, the latter corresponds to the Boltzmann-Gibbs distribution, which maximizes entropy in thermal equilibrium.  We observe that the empirical Lorenz curves move toward the calculated black exponential curve, approaching it closely from below.  \textcite{Lawrence-2013} attributed this behavior to globalization of the world economy, which mixes the world and brings it closer to the state of maximal entropy.  They predicted that global inequality will soon stop decreasing and will saturate at the Gini coefficient $G=0.5$ corresponding to the exponential distribution.  

This prediction was spectacularly confirmed in the follow-up paper by \textcite{Semieniuk-2020}, when the data for subsequent years up to 2017 became available.
Figure \ref{Fig:Gini} shows the Gini coefficient from 1980 to 2017 for CO$_2$ emissions per capita.  Black circles are the same data points for 1980--2010 as analyzed in \cite{Lawrence-2013}.  They manifest a decreasing trend and no sign of saturation yet.  In contrast, the new data points for 2011--2017 (red squares) exhibit saturation at the level of $G=0.5$, indicting that global inequality stopped decreasing soon after 2010.  This observation confirms the prediction made by \textcite{Lawrence-2013} on the basis of the maximal entropy principle.  Remarkably, this prediction was made at the time when global inequality has been steadily decreasing, and there was no indication of saturation in the available data yet.  The advanced prediction in \cite{Lawrence-2013} and the subsequent confirmation in \cite{Semieniuk-2020} strongly support the proposition that economic globalization is, indeed, governed by the principle of maximal entropy.

This observation has profound consequences for strategies and scenarios dealing with the climate change \cite{Semieniuk-2020}.  Various calls have been made for either lifting billions of people from poverty, which implies increased consumption and carbon emissions, or capping the level of per-capita CO$_2$ emissions at the top of the distribution above a certain threshold.  \textcite{Semieniuk-2020} constructed the Lorenz curves implied by these redistributive proposals and demonstrated that they would require an unprecedented reduction in global inequality, far below historical levels.  The recent saturation of the global inequality decrease further undermines feasibility of such redistributive scenarios.

A decreasing trend was also observed in global income inequality \cite{Milanovic-2012,Milanovic-2020,Milanovic-2023,Tharoor-2023,Salmon-2023}, 
but saturation has not been recognized yet.

\section{Conclusion}
\label{Sec:Conclusion}

The term ``econophysics'' was introduced in 1995 at a conference in Kolkata by the theoretical physicist Eugene Stanley for a new interdisciplinary research field applying methods of statistical physics to economics and finance \cite{Chakrabarti-2005}.   The paper by \textcite{Stanley-1996} presented a manifesto of the new field, arguing that ``behavior of large numbers of humans (as measured, e.g., by economic indices) might conform to analogs of the scaling laws that have proved useful in describing systems composed of large numbers of inanimate objects.''  About 30 years later, econophysics is now well recognized in physics \cite{Yakovenko-Rosser-2009} and is gradually penetrating into mainstream economics \cite{Shaikh-2017}.  Infusion of new ideas from a different field often results not in answering old questions within old framework, but in establishing new framework with new questions.  Much of misunderstanding and miscommunication between economists and physicists happens not because they are getting different answers, but because they are answering different questions \cite{Yakovenko-2022}.

The empirical demonstration of the two-class structure of income distribution, described by the simplest mathematical functions: exponential and power law, is a significant accomplishment of econophysics.  The Pareto power law for the upper tail has been known for a long time, but the exponential distribution, describing the majority of population, was not known before.  The ability to fit the whole income distribution by using only three parameters is somewhat reminiscent of Kepler's discovery that the vast amount of accumulated data for positions of planets can be reduces to a few parameters for their elliptic orbits.  Such data compression is a signature of scientific progress, which subsequently led Newton his discovery of the gravity law as an explanation for the elliptic orbits.  While social classes have been known in political economy since Karl Marx, realization that they are described by simple mathematical distributions is quite new.  As a follow-up to \cite{Dragulescu-2000}, \textcite{Wright-2005,Wright-2009} demonstrated emergence of two classes in more sophisticated agent-based simulations.  This work was further developed by \textcite{Isaac-2019} and in the book by \textcite{Cottrell-2009}, integrating economics, computer science, and physics.

The prediction \cite{Lawrence-2013} and confirmation \cite{Semieniuk-2020} of the saturation of the global inequality decrease is another significant accomplishment of econophysics.  As mentioned in the former paper, ``in physics, theories that not only explain known experiments, but also make successful predictions about future observations are particularly valuable.''  At the International Energy Workshop\footnote{https://www.internationalenergyworkshop.org/meetings-10.html} in 2017 at College Park, I chatted with the chief energy modeler of EIA.  He said in his talk that they have dynamical models, which are complicated and have many unknown parameters, so predictions cannot be made in practice.  I said that I do not have such models, but can make a valid prediction on the basis of a general principle (of entropy maximization).

There is discussion in the media about a global economic stagnation, sometimes called the ``economic ice age,'' pointing to various specific reasons \cite{Bott-2013}.  In contrast, \textcite{Lawrence-2013} suggested that the actual underlying reason is entropic.  Globalization was driving the world economy toward the state of maximal entropy, which has been achieved now, thus resulting in global stagnation.  Paradoxically, this advance toward the global statistical equilibrium may bring on social and geopolitical unrest because of slowdown in upward mobility for the lower part of global distribution and a downward slide from the previously privileged positions for the upper part \cite{Milanovic-2020,Milanovic-2023}.  On the other hand, the global economic stagnation may help to slow down the climate change.  The relentless growth, fueled by fossil fuels for the last couple centuries since the beginning of Industrial Revolution, must decelerate because of limited environmental capacity of the Earth.  However, the saturation of the global inequality decrease greatly complicates any agreement on limiting carbon emissions because of vast differences between the countries at the opposite ends of global distribution.  The urgently needed transition from redistributable hydrocarbon fuels toward decentralized locally-generated and locally-consumed renewable energy could help to both lower carbon emissions and reduce global inequality \cite{Lawrence-2013}.


\end{document}